\documentclass{PoS}

\usepackage{epstopdf}
\newcommand{\dEdx}{{\rm d}E/{\rm d}x}
\newcommand{\pt}{p_{\rm t}}

\newcommand{\Dzero}{{\rm D^0}}
\newcommand{\Dstar}{{\rm D^{*+}}}
\newcommand{\Dplus}{{\rm D^+}}

\newcommand{\raa}{R_{\rm AA}}
\newcommand{\rcp}{R_{\rm CP}}
\newcommand{\raah}{\raa^{{\rm h}^{\pm}}}
\newcommand{\raae}{\raa^{\rm electron}}
\newcommand{\raapi}{\raa^{\pi^{\pm}}}
\newcommand{\raaD}{R^{\rm D}_{\rm AA}}
\newcommand{\raaB}{R^{\rm B}_{\rm AA}}

\title{Open heavy flavour nuclear modification factor in ALICE}

\ShortTitle{Open heavy flavour $\raa$ in ALICE}

\author{\speaker{Andr\'e Mischke} for the ALICE Collaboration \\
       ERC-Research Group {\em QGP-ALICE}, Utrecht University\\
       Princetonplein 5, 3584 CC Utrecht, the Netherlands \\
       E-mail: \email{a.mischke@uu.nl}}

\abstract{The ALICE experiment has measured charm and beauty production in pp and Pb--Pb collisions at $\sqrt{s}$~= 2.76 and 7 TeV and $\sqrt{s_{\rm NN}}$~= 2.76 TeV, respectively, via the exclusive reconstruction of hadronic D meson decays and semi-leptonic D and B meson decays. In this contribution, recent results on the nuclear modification factor for open charmed mesons and heavy-flavour decay electrons at central rapidity and muons at forward rapidity are presented and discussed.}

\FullConference{50th International Winter Meeting on Nuclear Physics \\
		          23-27 January 2012 \\
		          Bormio, Italy}

\begin{document}

%
\section{Introduction}
High-energy collisions of heavy atomic nuclei allow to explore the behaviour of strongly interacting matter at high temperatures, where a new phase of matter, the Quark-Gluon Plasma (QGP), is predicted to exist (for a recent review see~\cite{intro}). 
Heavy quarks (charm and beauty), abundantly produced in heavy ion collisions at the Large Hadron Collider (LHC), provide sensitive penetrating probes of such matter. Due to their large mass, they are believed to be predominantly produced in the initial state of the collision by gluon fusion processes and thus provide information about the hottest initial phase. These heavy quarks propagate through the hot and dense QCD matter and lose energy through medium-indiuced gluon radiation (colour charge dependent) and collisions with the medium. Moreover, theoretical models predicted that heavy quarks should experience smaller energy loss than light quarks due to the suppression of small angle gluon radiation ({\em dead-cone effect}~\cite{deadcone1, deadcone2}).
Thus, the study of heavy-flavour production in nucleus-nucleus collisions provides key tests of parton energy-loss models yielding profound insight into the properties of the produced QCD matter.

Medium effects are typically quantified using the nuclear modification factor $\raa$ where the particle yield in Pb--Pb collisions is divided by the yield in pp reactions scaled by the number of binary collisions. $\raa$=1 would indicate that no nuclear effects, such as Cronin effect, shadowing or gluon saturation, are present and that nucleus-nucleus collisions can be considered as an incoherent superposition of nucleon-nucleon interactions.
By comparing the nuclear modification factor of charged pions $(\raapi$), mostly originating from gluon fragmentation at this collision energy, with that of hadrons with charm $\raaD$ and beauty $\raaB$ the dependence of the energy loss on the parton nature (quark/gluon) and mass can be investigated.
A mass ordering pattern $\raapi<\raaD<\raaB$ is expected~\cite{theo:wicks2007}.

%
\section{ALICE detector setup, trigger and data set}
\begin{figure}[!t]
  \begin{center}
  \includegraphics[width=0.44\textwidth]{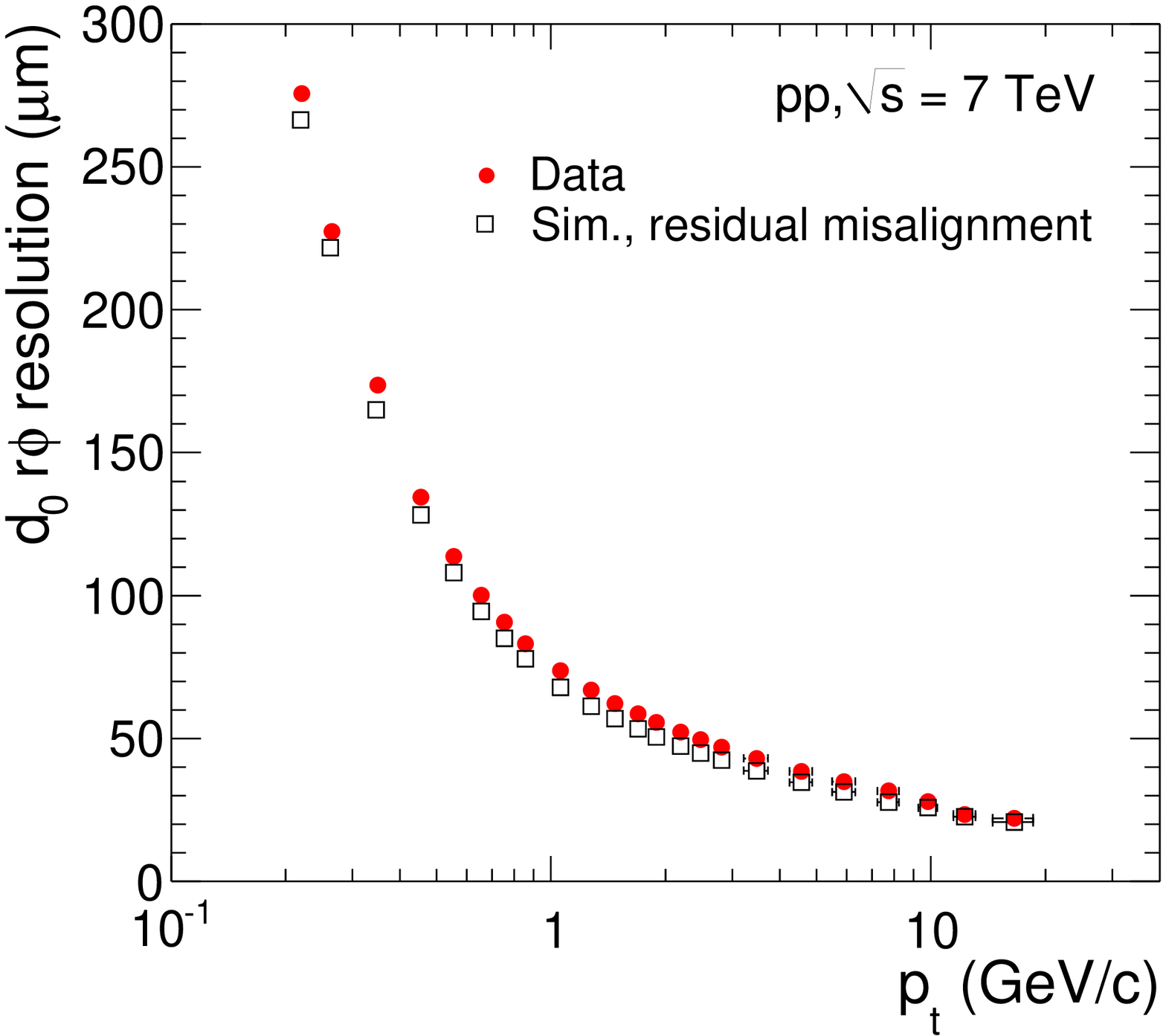}
  \hspace{6mm}
  \includegraphics[width=0.47\textwidth]{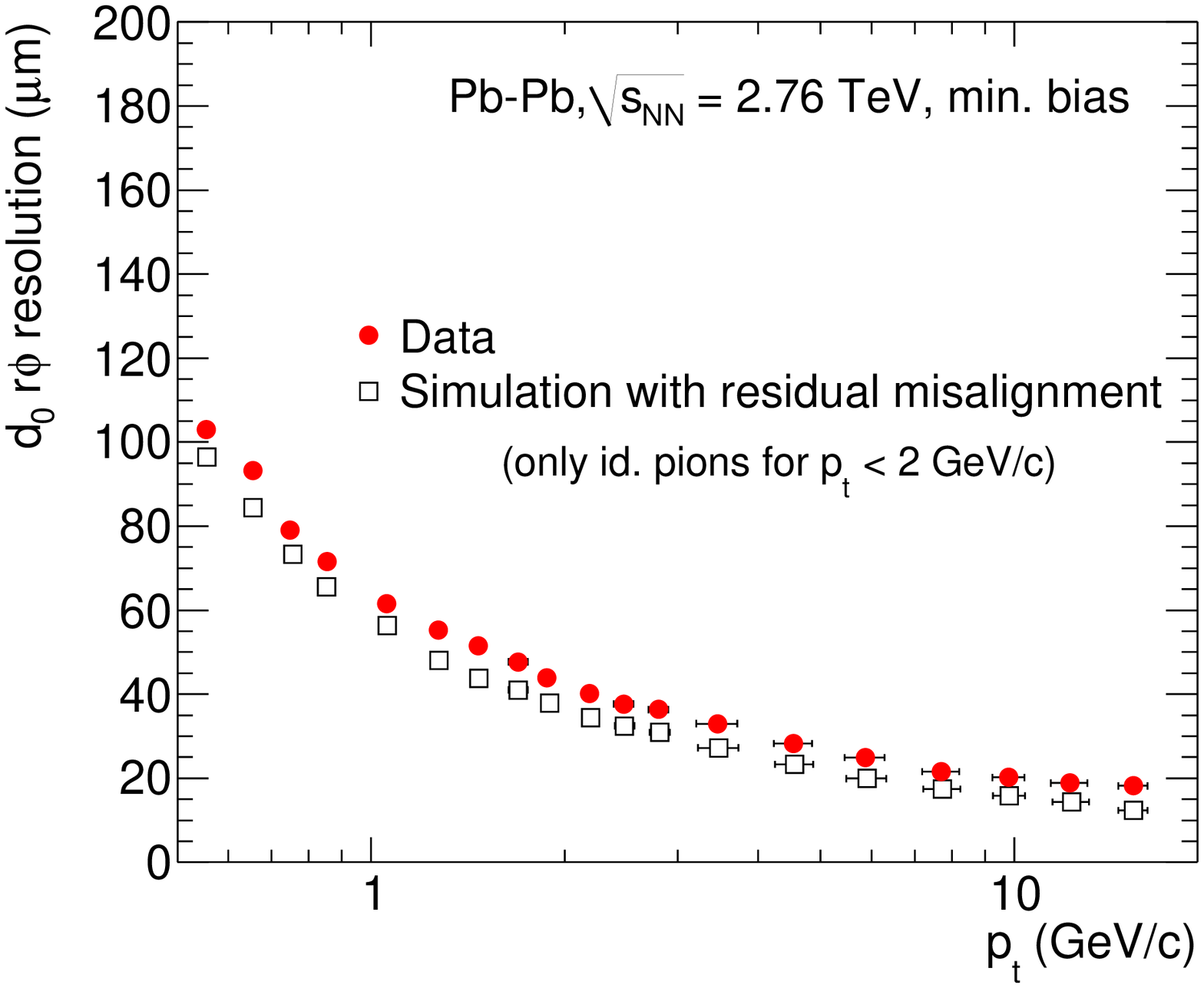}
  \vspace{-3mm}
  \caption{Track impact parameter ($d_0$) resolution in the transverse plane ($r\phi$ direction) as a function of $\pt$ in pp (left panel) and Pb-Pb collisions (right panel) comparing data and Monte Carlo simulation~\cite{alice:itsalign}. This resolution includes the uncertainty in the primary vertex position, which is reconstructed excluding the track being probed. For $\pt < 2$ GeV/$c$, pion identification by the TPC or TOF detectors is required.}
\label{Fig:1}
\end{center}
\end{figure}
ALICE, A Large Ion Collider Experiment, is the dedicated detector for measurements on high energy heavy-ion collisions~\cite{alice:PPR2, alice:detpaper}. Results with the first LHC Pb--Pb data are reported in~\cite{alice:first1, alice:first2}.
Its characteristic features are the very low momentum cut-off (100 MeV/$c$), the low material budget and the excellent particle identification and vertexing capabilities in a high multiplicity environment.
At mid-rapidity, tracking with up to 159 three-dimensional space-points and particle identification through the measurement of the specific ionisation energy loss (d$E$/d$x$) is performed using the large volume Time Projection Chamber (TPC), located inside the large solenoidal magnet with a field of $B$~= 0.5 T. The TPC has a coverage from -0.9 to 0.9 in pseudo-rapidity and 2$\pi$ in azimuth. The relative d$E$/d$x$ resolution is 5-6$\%$ and the relative momentum resolution $\Delta p/p$ is 5$\%$ at 100 GeV/$c$.
The particle identification up to about $\pt <$ 2 GeV/$c$ is performed with the Time Of Flight (TOF) detector, which has an intrinsic time resolution better than 100~ps.
The Inner Tracking System consists of six concentric cylindrical layers of silicon detectors and provides excellent reconstruction of displaced vertices with a transverse impact parameter resolution better than 65 $\mu$m for 
$\pt >$ 1 GeV/$c$~\cite{alice:itsalign}.
The track impact parameter resolution in the transverse plane in pp and Pb--Pb collisions is depicted in Fig.~\ref{Fig:1}.
The total material budget for radial tracks in the transverse plane is $\approx 7.7\%$ of the radiation length.
%
%
%
At forward rapidity ($-4<\eta<-2.5$) muons are identified in the muon spectrometer. The muon spectrometer detects muons with momentum larger than 4 GeV/$c$ and is composed of two absorbers, a dipole magnet providing a field integral of 3 Tm, and tracking and trigger chambers. Tracking is performed by means of five tracking chambers with an intrinsic spatial resolution better than 100 $\mu$m.

The Pb--Pb data presented in this paper were collected with a minimum-bias trigger in November-December 2010 during the first run with heavy-ions at the LHC at a centre-of-mass energy of $\sqrt{s_{\rm NN}}$~= 2.76 TeV per nucleon-nucleon pair.
This trigger is based on the information of the two innermost silicon layers ($|\eta| < 2$) and the VZERO scintillator hodoscopes ($2.8<\eta<5.1$ and $-3.7<\eta<-1.7$). The efficiency for triggering hadronic interactions was 100\% for Pb--Pb collisions in the centrality range considered in the analyses. Beam background collisions were removed offline on the basis of the timing information provided by the VZERO and the Zero Degree Calorimeters (located near the beam pipe at $z\pm 114$ m from the interaction point). Only events with a vertex position within 10 cm from the centre of the detector along the beam line were used, resulting in 17M events. The forward single-muon arm triggers in coincidence with the minimum-bias trigger.
The Pb--Pb data are classified based on their centrality defined in terms of percentiles of the hadronic Pb--Pb cross section and determined from the distribution of the summed amplitudes in the VZERO detector. This distribution was fitted using the Glauber model for the geometrical description of the nuclear collision together with a two-component model for particle production~\cite{firstApaper}.

The pp results correspond to 315M and 180M minimum-bias events at $\sqrt{s}$~= 7 TeV for the D meson (integrated luminosity: 5~nb$^{-1}$) and single electron analyses (2.6~nb$^{-1}$), respectively, and 2M muon triggered events (16.5~nb$^{-1}$).

%
\section{Results}
The following sections give a brief description of the analyses and results on charm and beauty production in pp and Pb-Pb collisions at $\sqrt{s}$~= 2.76 and 7 TeV and $\sqrt{s_{\rm NN}}$~= 2.76 TeV, respectively, via the exclusive reconstruction of hadronic and semi-leptonic decays of D and B mesons. Some of those results are already available as journal publications: 
D meson production in pp reactions at $\sqrt{s}$~= 7 TeV~\cite{7TeVD} and in Pb--Pb collisions at $\sqrt{s_{\rm NN}}$~= 2.76 TeV~\cite{RaaD} and heavy-flavour decays muons in pp interactions at $\sqrt{s}$~= 7 TeV~\cite{mu7TeV}.

%
\subsection{D mesons at mid-rapidity}
\begin{figure}[!t]  
\begin{center}        
  \includegraphics[width=0.3294\textwidth]{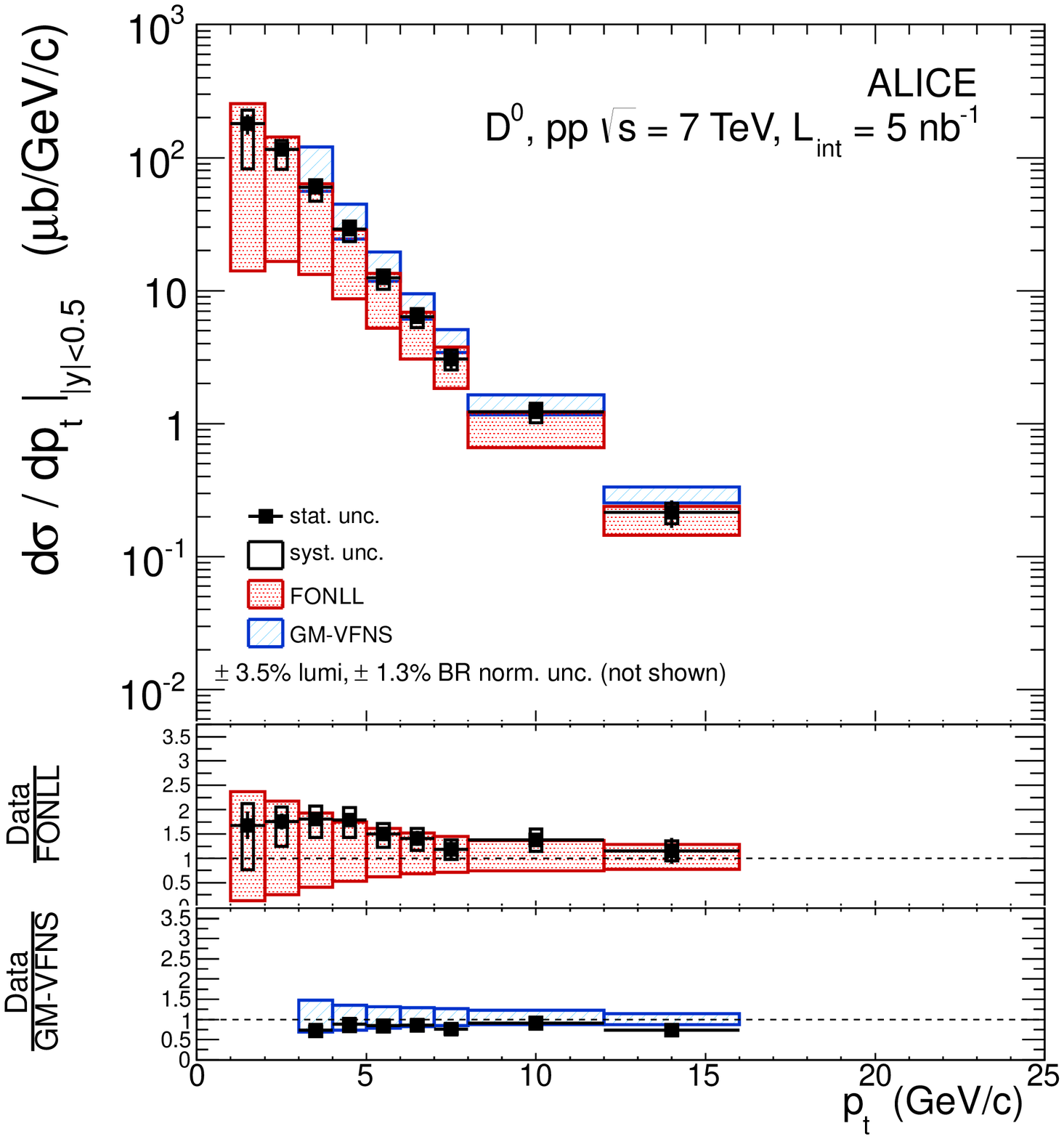}
  \includegraphics[width=0.3294\textwidth]{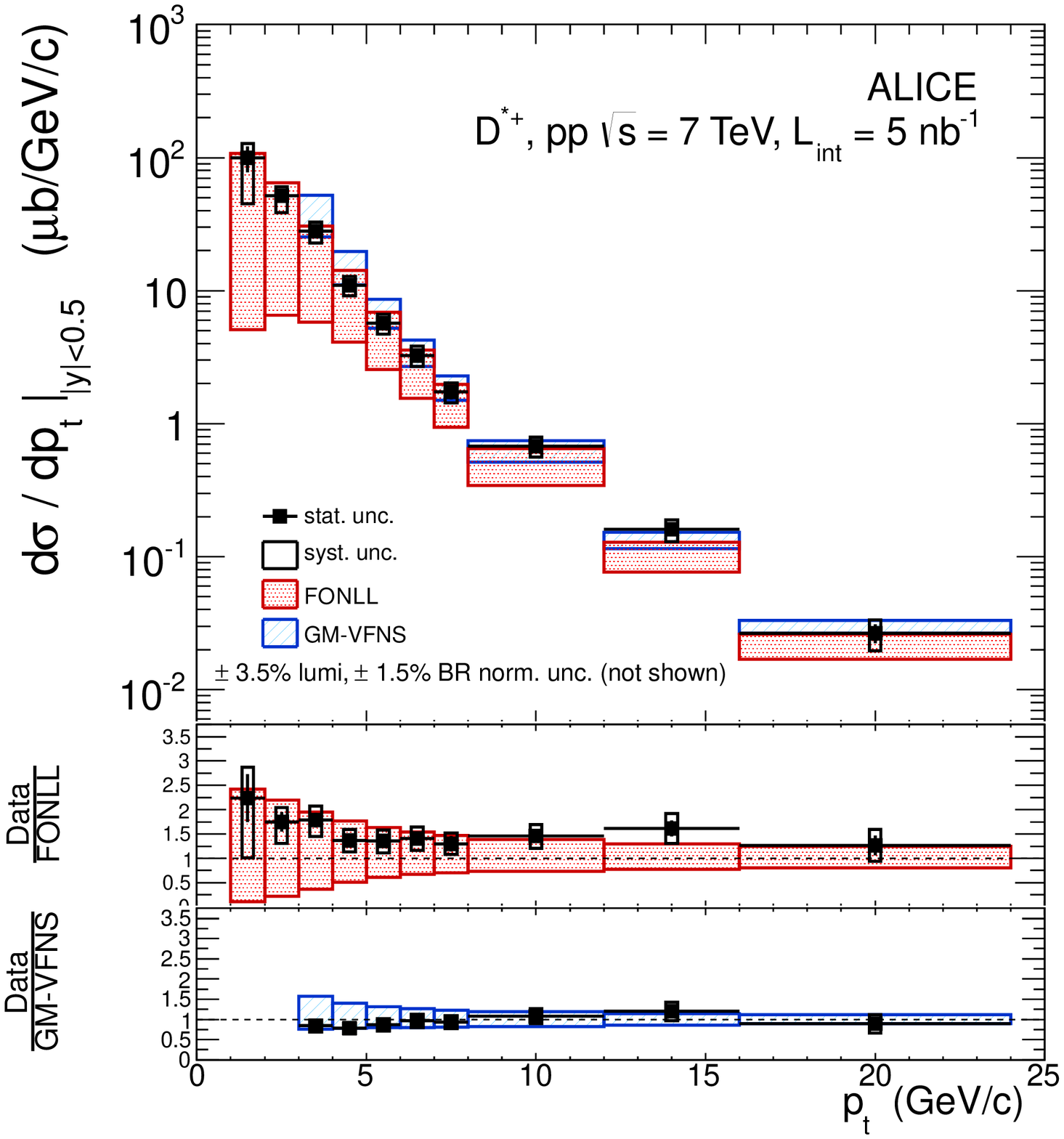}
  \includegraphics[width=0.3294\textwidth]{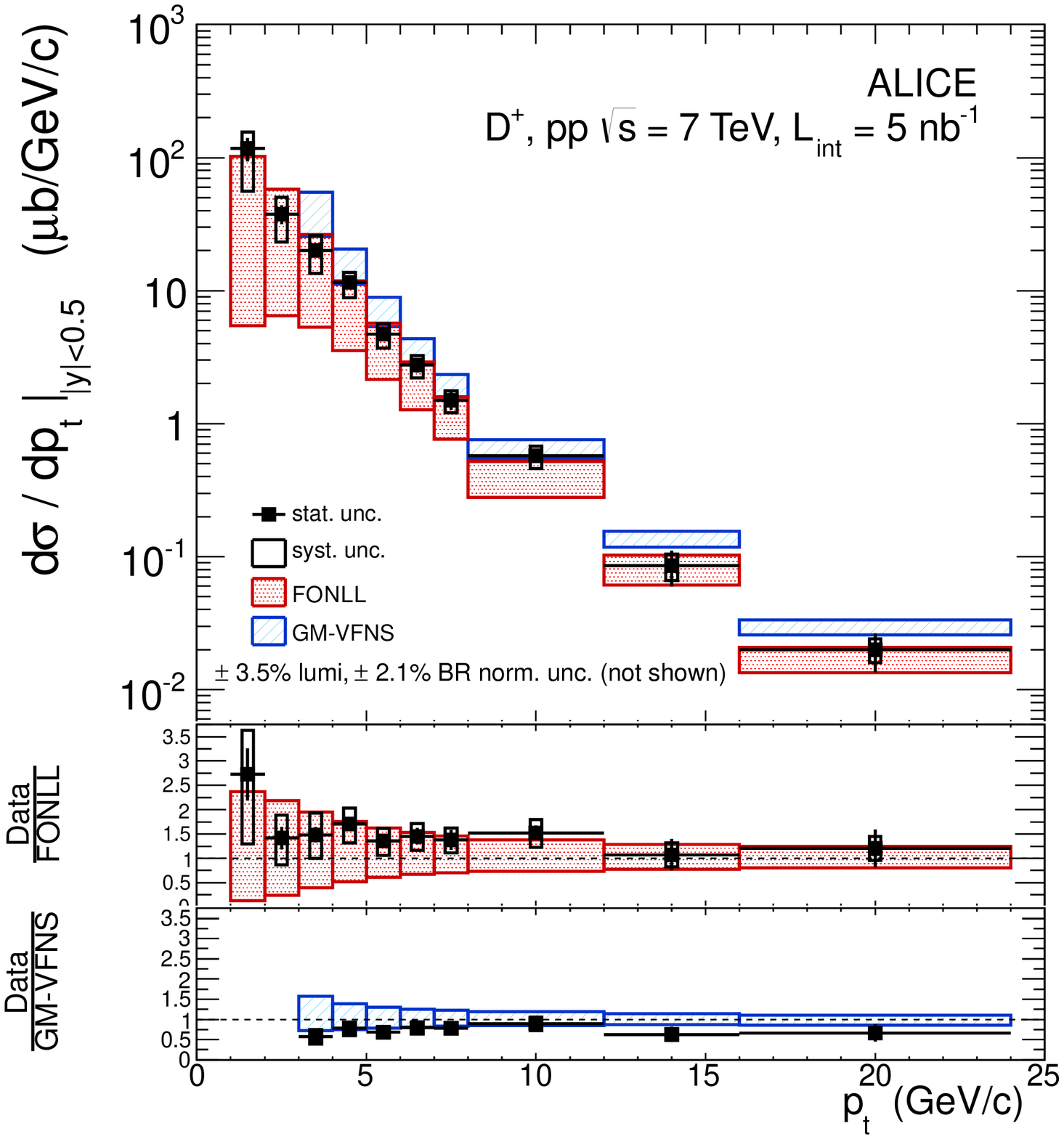}
  \caption{(colour online) $\pt$-differential inclusive cross section for prompt $\Dzero$, $\Dstar$ and $\Dplus$ mesons for $|\eta|<0.5$ in pp collisions at $\sqrt{s}$~= 7 TeV~\cite{7TeVD}, compared with FONLL~\cite{theo:fonll} and GM-VFNS~\cite{theo:vfns} theoretical calculations. The symbols are positioned horizontally at the centre of each $\pt$ interval. The normalization uncertainty is not shown (3.5\% from the minimum-bias cross section plus the branching ratio uncertainties).}
\label{Fig:21}
\end{center}
\end{figure}
D$^{0}$, D$^{*+}$ and D$^+$ mesons are reconstructed through their hadronic channels: 
D$^0\rightarrow$ K$^- \pi^+$ (BR = ($3.91\pm0.05)\%$), 
D$^{*+}\rightarrow$ D$^0\pi^+$ (BR = ($67.7\pm 0.5)\%$) and
D$^+\rightarrow$ K$^+\pi^-\pi^+$ (BR = $(9.22\pm 0.21)\%$), 
based on their decay topology and the invariant mass technique. The kaon and pion identification using TPC and TOF helps to reduce background at low $\pt$. Details of the analysis can be found in~\cite{7TeVD, RaaD}.
The feed-down from beauty decays is calculated from theory and gives a contribution of 10-15\%.
The differential cross section for prompt D$^0$, D$^{*+}$ and D$^+$ mesons, measured in the $\pt$ range 1--24 GeV/$c$, is shown in Fig.~\ref{Fig:21}. The data are well described within uncertainties by perturbative QCD calculations at Fixed-Order plus Next-to-Leading Logarithm (FONLL) level~\cite{theo:fonll} and GM-VFNS predictions~\cite{theo:vfns}.

\begin{figure}[t]
  \centering
  \includegraphics[scale=0.32]{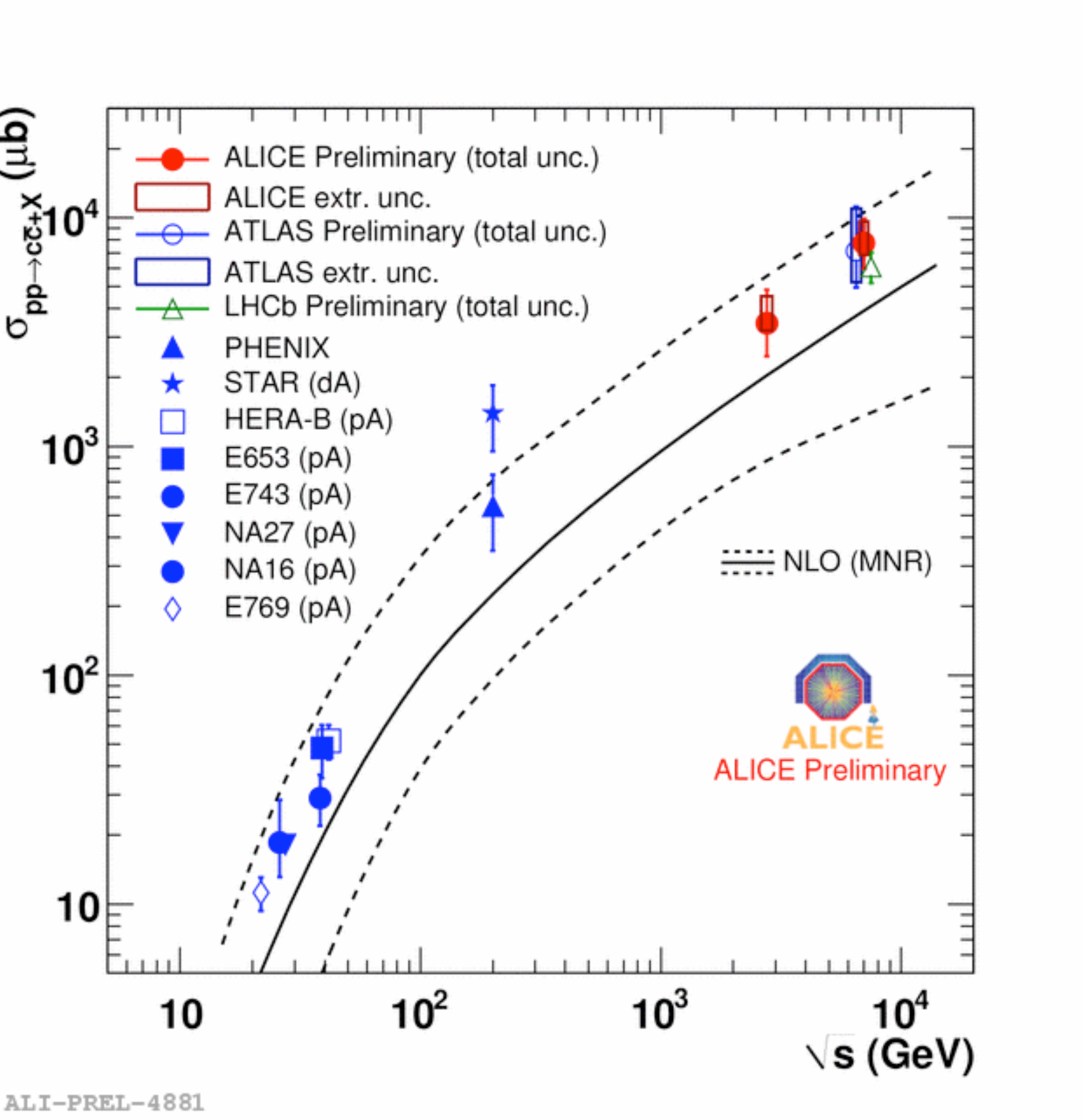}
  \vspace*{-0.4cm}
  \caption{Comparison of total charm cross section in pp collisions at $\sqrt{s}$~= 2.76 and 7 TeV with the world data. The NLO MNR calculation~\cite{mnr} (and its uncertainties) is represented by solid (dashed) lines.}  
\label{Fig:22}
\end{figure}

The D meson production cross section was also measured in pp collisions at $\sqrt{s}$~= 2.76 TeV 
($L_{int}$ = 1.1 nb$^{-1}$).
The total charm cross section was extracted by extrapolating the measurements to the full phase space. 
The measured charm cross sections fit into the world data trend and is well described by NLO MNR calculation~\cite{mnr} as shown in Fig.~\ref{Fig:22}.
These data are important baseline measurements for quarkonia studies in Pb-Pb collisions and the corresponding parton spectra from pQCD serve as an input for energy loss models.

%
\begin{figure}[!t]
  \begin{center}
  \includegraphics[width=0.80\textwidth]{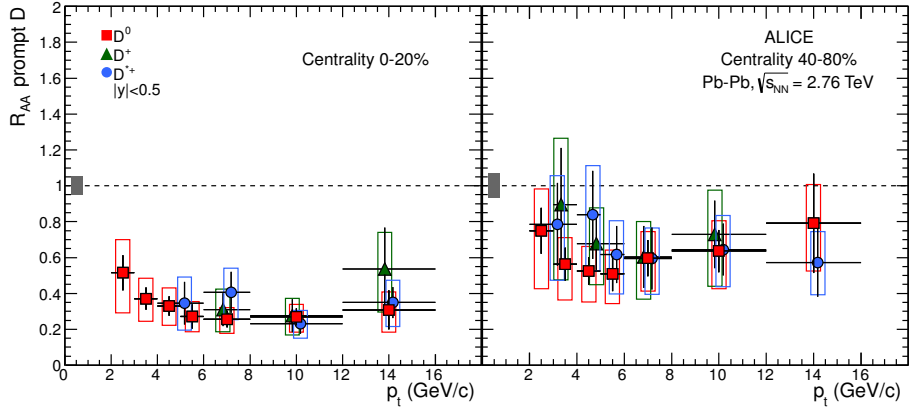}
  \vspace{-2.5mm}
  \caption{(colour online) $\raa$ for prompt D$^0$, D$^{*+}$ and D$^+$ in the 0--20\% (left) and 40--80\% (right) centrality classes~\cite{RaaD}. Statistical (bars), systematic (empty boxes), and normalization (full box) uncertainties are shown. Horizontal error bars reflect bin widths, symbols were placed at the centre of the bin.}
\label{fig:RAApt}
\end{center}
\end{figure}
\begin{figure}[!t]
  \begin{center}
  \includegraphics[width=0.40\textwidth]{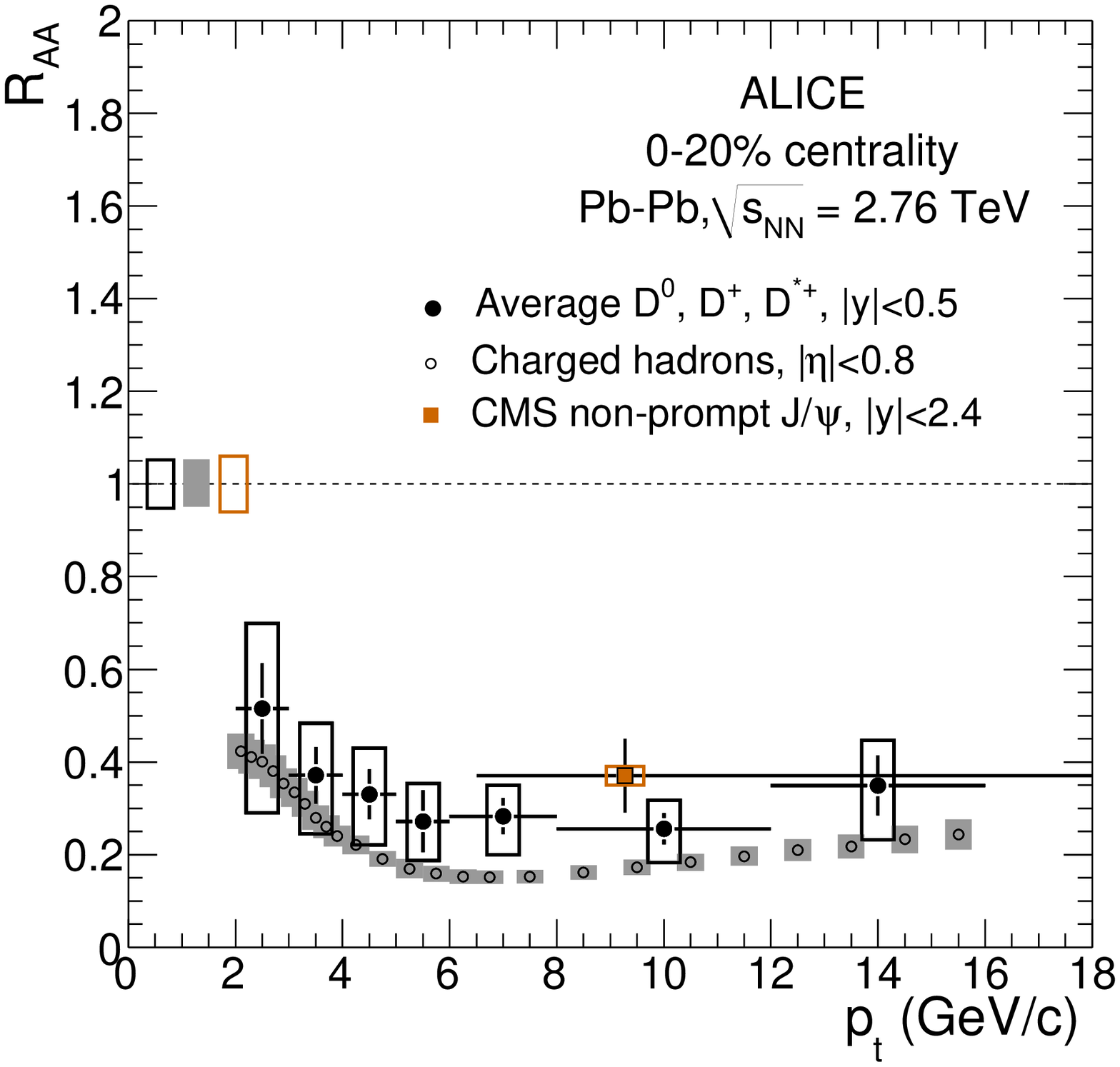}
  \hspace{7mm}
  \includegraphics[width=0.40\textwidth]{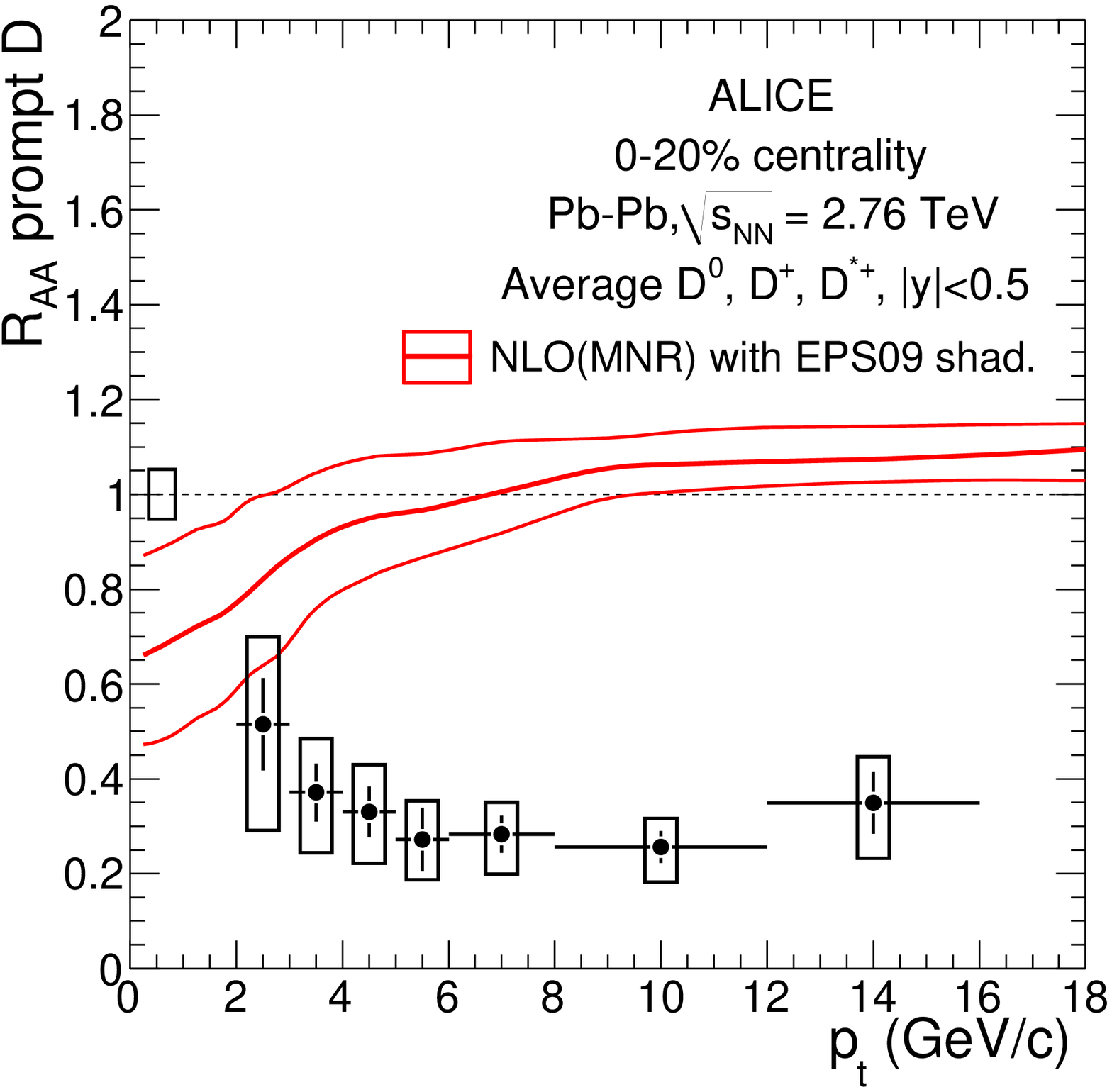}
  \vspace{-4mm}
  \caption{Average $\raa$ of D mesons in the 0--20\% centrality class~\cite{RaaD} compared to: 
  left, the nuclear modi\-fication factors of charged hadrons~\cite{pionRaa} in the same centrality class
  and non-prompt $\rm J/\psi$ from B decays; 
  right, the expectation from NLO pQCD~\cite{mnr} with nuclear shadowing~\cite{eps09}.}
\label{fig:RAApt_eps_charged}
\end{center}
\end{figure}

The first measurement of the nuclear modification factor of prompt D$^0$, D$^{*+}$ and D$^+$ mesons in Pb--Pb collisions at $\sqrt{s_{\rm NN}}$~= 2.76 TeV is shown in Fig.~\ref{fig:RAApt} for the centrality classes 0--20\% and 40--80\%. The $\raa$ of the three D meson species agree within statistical uncertainties and shows for the 20\% most central collisions a strong suppression by a factor of 4-5 for $\pt >$ 5 GeV/$c$. 
Less suppression is observed for the centrality class 40--80\%.
The average $\raa$ of D mesons in the 0--20\% centrality class is compatible, within uncertainties, with the $\raa$ of charged hadrons~\cite{pionRaa} (cf. Fig.~\ref{fig:RAApt_eps_charged}, left panel).
There are indications that $\raaD>\raah$ at lower $\pt$, although more data are necessary to draw final conclusions.
In Fig.~\ref{fig:RAApt_eps_charged} (right panel) the $\raaD$ measurement is compared with expectation from the NLO pQCD calculations~\cite{mnr} with nuclear shadowing using the EPS09 parametrisation~\cite{eps09}.  
Only little shadowing is expected in this $\pt$ range. Thus the observed suppression is a final state effect, namely due to the interaction of the heavy quarks with the hot QCD medium.
\begin{figure}[!t]
  \vspace{-4mm}
  \begin{center}
\includegraphics[width=0.7\textwidth]{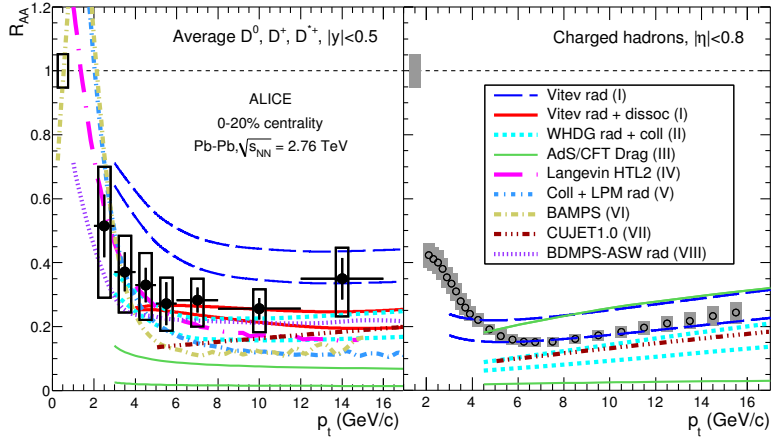}
  \vspace{-4mm}
  \caption{(colour online) Average $\raa$ of D mesons~\cite{RaaD} (left) and $\raa$ of charged hadrons~\cite{pionRaa} (right) in the 0--20\% centrality class compared to different model calculations based on parton energy loss (see~\cite{RaaD} for details). The two normalisation uncertainties are almost fully correlated.}
\label{fig:RAApt_models}
\end{center}
\end{figure}
In Fig.~\ref{fig:RAApt_models} the nuclear modification factor for the average D mesons and charged hadrons in the 20\% most central Pb-Pb collisions are confronted with several theoretical models based on parton energy loss (see~\cite{RaaD} for details). 
Among the models that compute both observables, radiative energy loss supplemented with in-medium D meson dissociation and radiative plus collisional energy loss in the WHDG and CUJET1.0 implementations simultaneously describe the charm and light-flavour suppression reasonably well.
While in the former calculation the medium density is tuned to describe the inclusive jet suppression at LHC energies, for the latter two it is extrapolated to LHC conditions starting from the value that describes the pion suppression at RHIC energy. This might explain why these two models are somewhat low with respect to the charged-hadron $\raa$. A model based on AdS/CFT drag coefficients underestimates significantly the charm $\raa$ and has very limited predictive power for the light-flavour $\raa$.

%
\subsection{Single electrons from heavy-flavour decays at mid-rapidity}
\begin{figure}[t]
  \centering
  \includegraphics[scale=0.35]{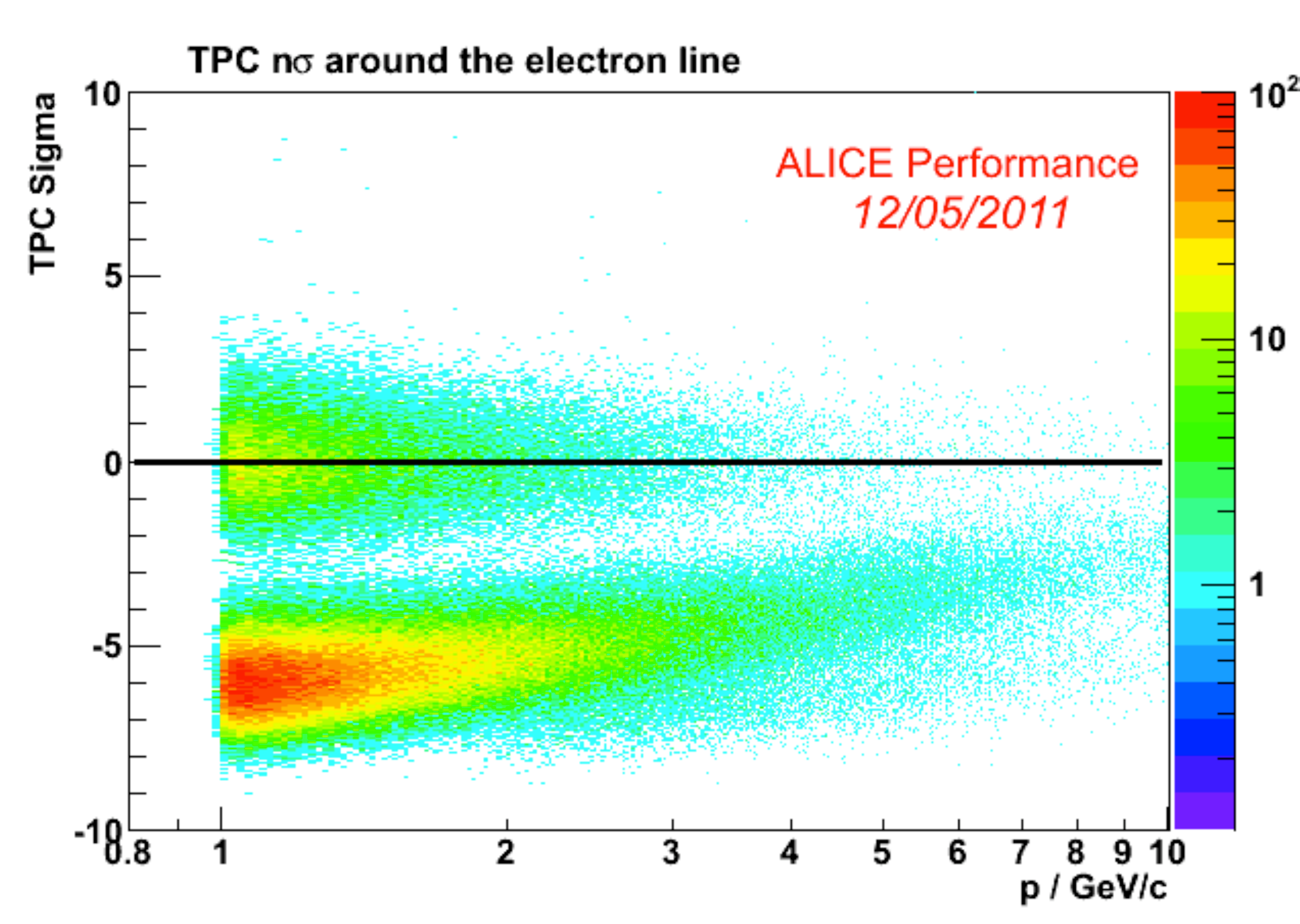}
  \vspace*{-4mm}
  \caption{Electron identification using the normalized TPC $\dEdx$ distribution. The upper band is electrons, the lower one charged pions.}  
\label{Fig:31}
\end{figure}
\begin{figure}[t]
  \centering
  \hspace{2mm}
  \includegraphics[width=0.51\textwidth]{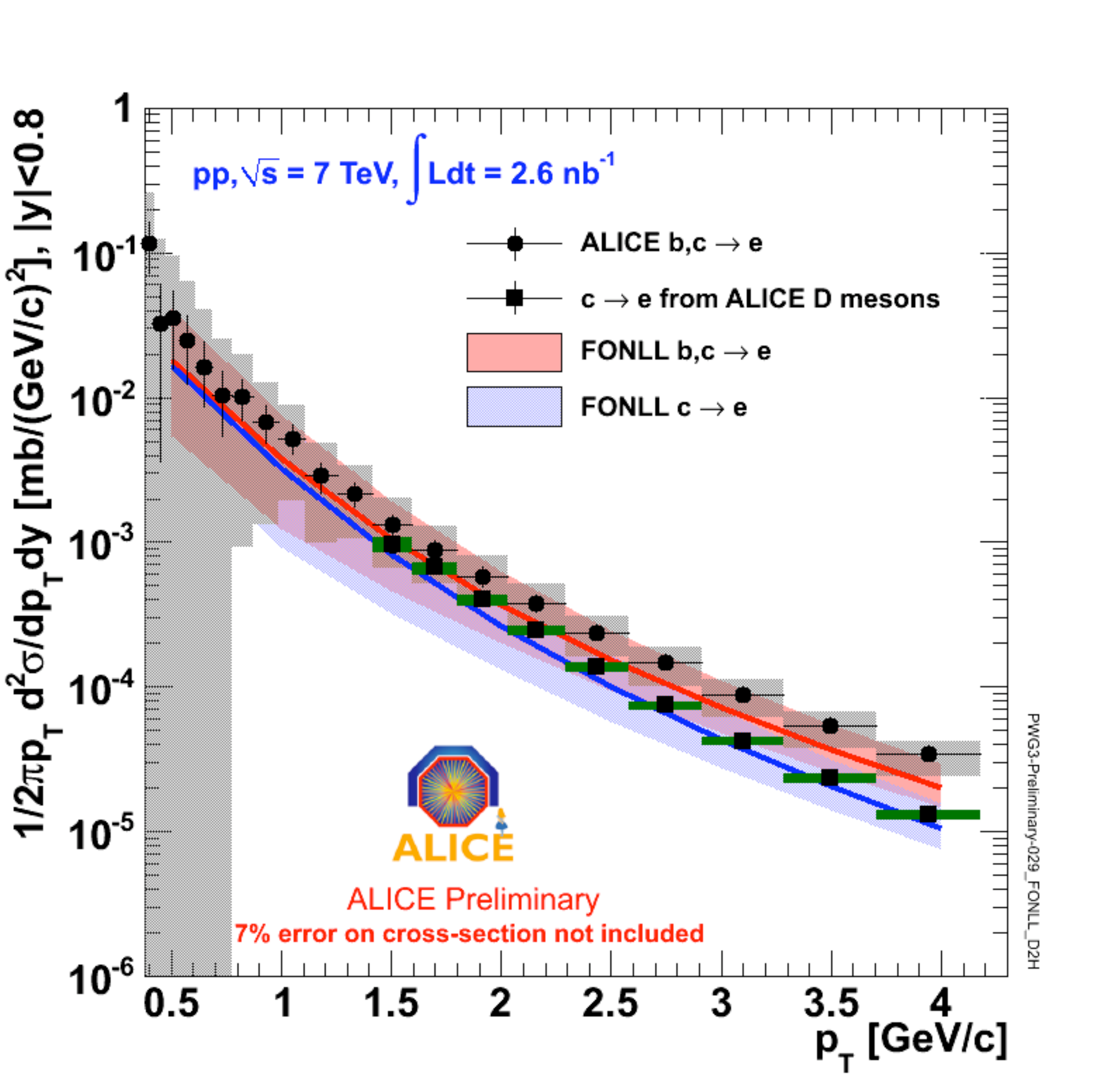}
  \hspace{3mm}
  \includegraphics[scale=0.33]{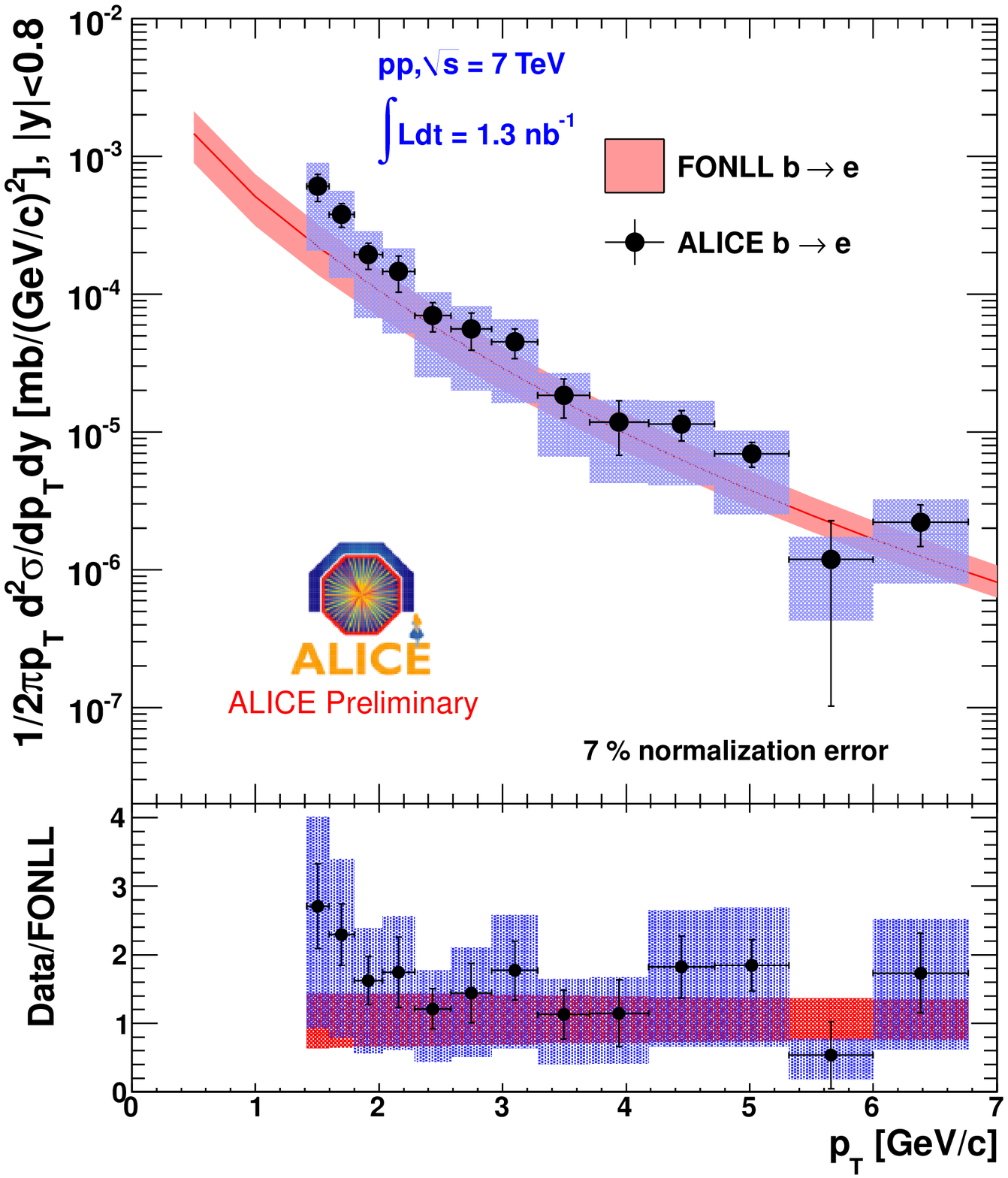}
  \vspace*{-9mm}
  \caption{Invariant cross section of electrons from charm or beauty decays (left panel) and beauty decays only (right panel) in pp collisions at $\sqrt{s}$~= 7 TeV, compared to FONLL calculations (red and blues curves) and to expectations from D meson decay electrons (left panel).}  
\label{Fig:32}
\end{figure}
The  cross section for open heavy-flavour production at mid-rapidity has also been studied through the measurement of single electrons. Those electrons are identified, using the TPC and TOF detectors, on a statistical basis by subtracting a cocktail of background electrons from the inclusive electron spectrum. This background arises mainly from $\gamma$ conversion electrons in the detector material and $\pi^0$ Dalitz decays. For $\pt$ up to a few GeV/$c$ this cocktail can be determined precisely by means of the measured $\pi^0$ cross section.
Figure~\ref{Fig:31} depicts the momentum dependence of the normalized TPC d$E$/d$x$ distribution after applying a cut on the TOF signal, which rejects kaons ($<$ 1.5 GeV/$c$) and protons ($<$ 3 GeV/$c$). A d$E$/d$x$ cut separates electrons from charged pions up to about 10 GeV/$c$ with a residual pion contamination of less than 15\%.
Figure~\ref{Fig:32} (left panel) illustrates the single electron cross section in pp reactions at $\sqrt{s}$~= 7 TeV, which has a total systematic uncertainty of 16-20\% ($\pt$ dependent) plus 7\% for the normalisation. The data are well described by FONLL calculations~\cite{theo:fonll} within uncertainties. Moreover, the low $\pt$ single electron spectrum agrees with expectations from D meson decay electrons. Similar measurements at high-$\pt$ using the EMCal and TRD detectors are ongoing.

Electrons from beauty decays are identified through displaced vertices. 
A selection is applied by requiring electrons to miss the interaction vertex by at least three times the error on the separation of the track from the vertex (of the order of 80$\mu$m at 2 GeV/$c$). The remaining contribution from charm
decays is estimated from the D meson cross section at mid-rapidity and subtracted.
The resulting spectrum of electrons from beauty decays is shown in Fig.~\ref{Fig:32} (right panel) and is well described by FONLL calculations~\cite{theo:fonll} within uncertainties.

\begin{figure}[t]
  \centering
  \includegraphics[width=0.47\textwidth]{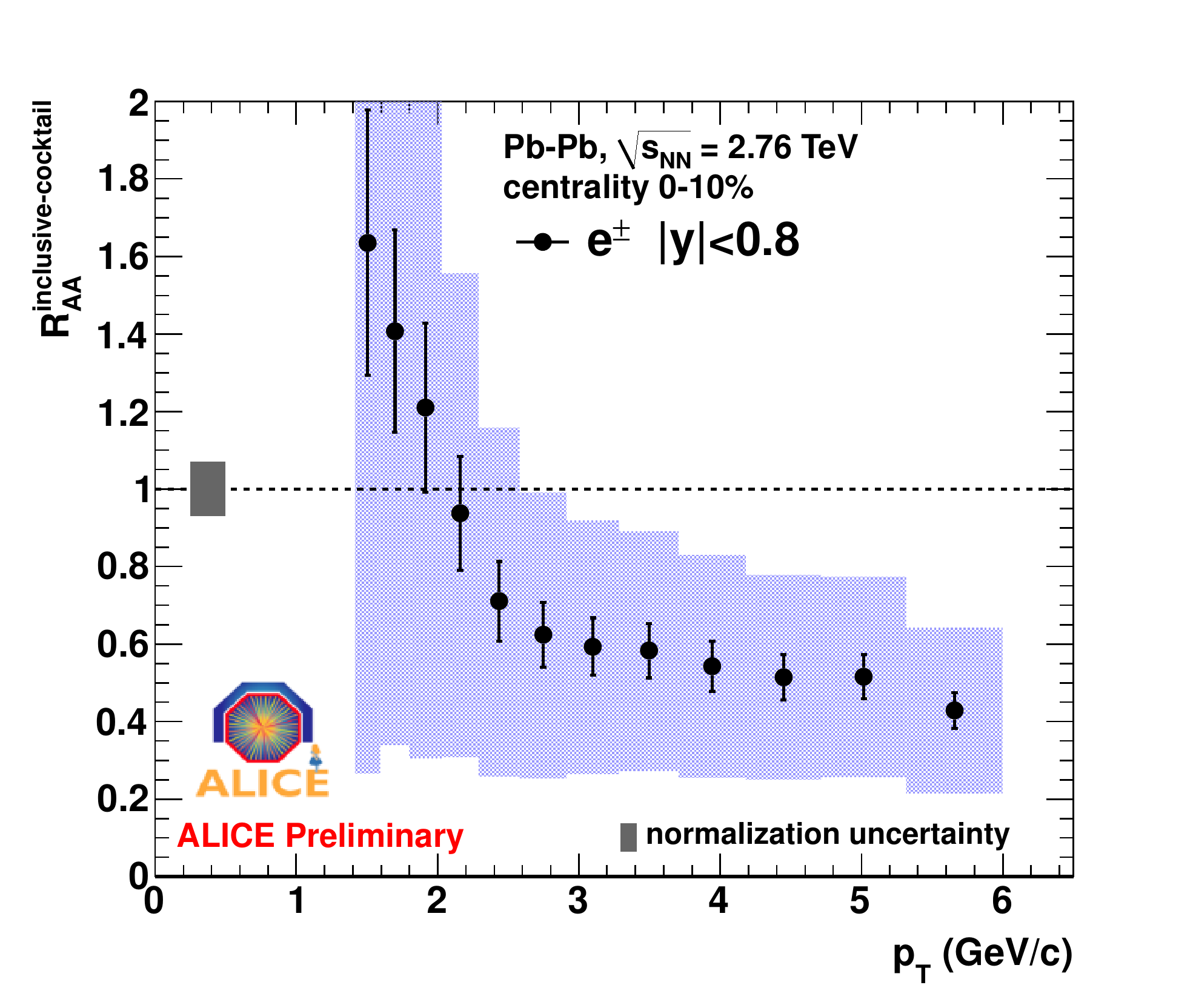}
  \vspace*{-0.4cm}
  \caption{Nuclear modification factor for cocktail-subtracted electrons in the 10\% most central Pb-Pb collisions at $\sqrt{s_{\rm NN}}$~= 2.76 TeV. The blue boxes are the systematic uncertainties.}
\label{Fig:33}
\end{figure}

%
A similar analysis was performed with the 2010 Pb--Pb data set. 
The hadron contamination in the electron sample remains below 10\% in the momentum range between 1.5 and 6 GeV/$c$.
The nuclear modification factor of cocktail-subtracted electrons for the most 10\% central Pb--Pb collisions at $\sqrt{s_{\rm NN}}$~= 2.76 TeV is shown in Fig.~\ref{Fig:33}. The $\raae$ and $\raaD$ at high $\pt$ (cf. Fig.~\ref{fig:RAApt}) are consistent within uncertainties.
Studies are ongoing to determine the bottom contribution to the single electrons experimentally using the electron displacement method and azimuthal angular correlations~\cite{relB}.
However, FONLL calculations indicate that b-hadron decays start to dominate above $\approx$5-6 GeV/$c$. Thus, beauty suppression appears to be large.

%
\subsection{Single muons from heavy-flavour decays at forward rapidity}
\begin{figure}[t]
  \centering
  \includegraphics[width=0.7\textwidth]{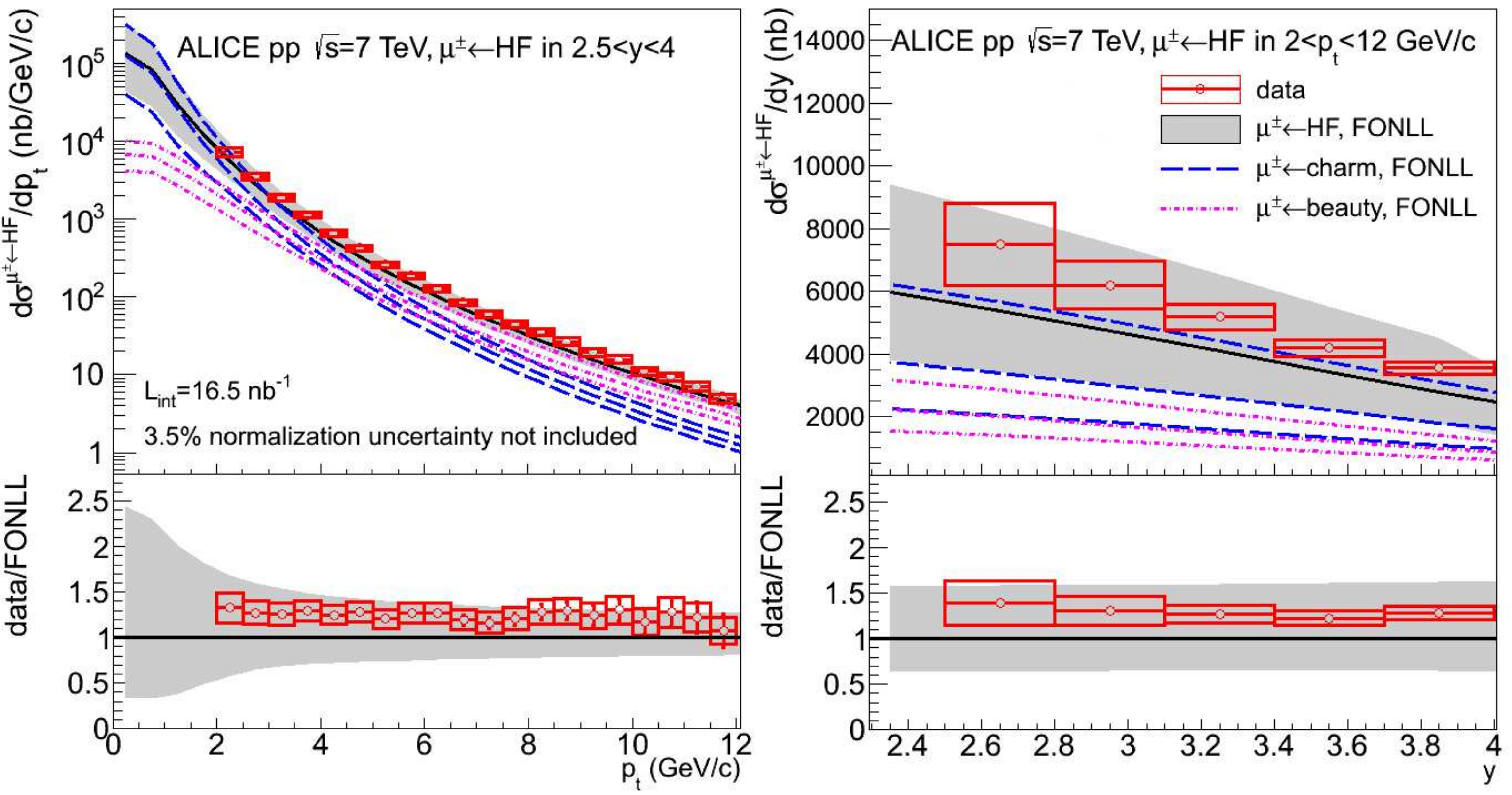}
  \vspace*{-0.3cm}
  \caption{Differential transverse momentum cross section for single muons in -4$<\eta<$-2.5~\cite{mu7TeV}. The statistical error is smaller than the markers. The systematic errors (open boxes) do not include an additional 10\% error on the minimum-bias pp cross section. The grey band indicates the FONLL prediction.}
\label{Fig:41}
\end{figure}

Heavy-flavour production at forward rapidities ($-4<\eta<-2.5$) is studied with single muons using the ALICE muon spectrometer. The extraction of the heavy-flavour contribution of the single muon spectra requires the subtraction of three main background sources: 
a) muons from the decay-in-flight of light hadrons (decay muons); 
b) muons from the decay of hadrons produced in the interaction with the front absorber (secondary muons); 
c) punch-through hadrons.
The last contribution can be efficiently rejected by requiring the matching of the reconstructed tracks with the tracks in the trigger system. 
Due to the lower mass of the parent particles, the background muons have a softer $\pt$ distribution than the heavy-flavour muons, and dominate the low-$\pt$ region. Therefore, the analysis is restricted to the $\pt$ range 2-12 GeV/$c$. 
Simulation studies indicate that, in this $\pt$ range, the contribution of secondary muons is small (about 3\%). 
The main source of background in this region consists of decay muons (about 25\%), which have been subtracted using Monte Carlo simulations. 
Figure~\ref{Fig:41} illustrates the cross section of heavy-flavour decay muons as a function of $\pt$ (left panel) and $y$ (right panel) in pp collisions at $\sqrt{s}$~= 7 TeV~\cite{mu7TeV}. The FONLL calculation~\cite{theo:fonll} agrees with the data within uncertainties. 
\begin{figure}[t]
  \centering
  \includegraphics[width=0.5\textwidth]{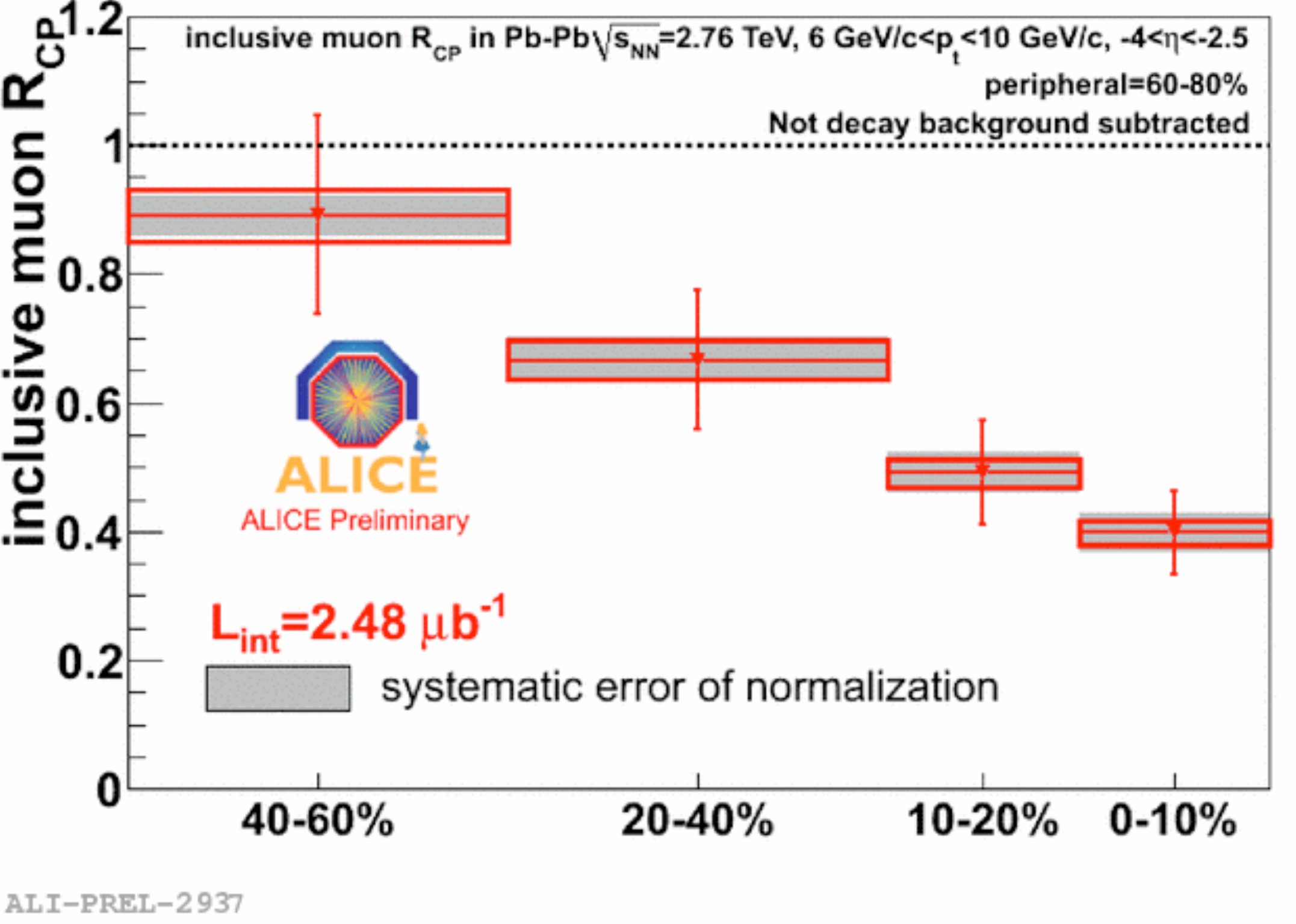}
  \vspace*{-0.4cm}
  \caption{Centrality dependence of the $\rcp$ for single muons in the $\pt$ range 6-10 GeV/$c$ at forward rapidity in Pb--Pb collisions at $\sqrt{s_{\rm NN}}$~= 2.76 TeV.}  
\label{Fig:42}
\end{figure}
Single muons were measured in Pb--Pb collisions at $\sqrt{s_{\rm NN}}$~= 2.76 TeV using a similar analysis procedure.
The $\rcp$ for single muons in the $\pt$ range 6-10 GeV/$c$ shows a clear centrality dependence (cf. Figure~\ref{Fig:42}).

\section{Summary}
Heavy quarks are particularly good probes to study the properties of hot QCD matter (especially the transport
properties).
The $\raa$ of D$^0$, D$^{*+}$ and D$^+$ was measured for the first time as a function of transverse momentum and centrality in Pb--Pb collisions at $\sqrt{s_{\rm NN}}$~= 2.76 TeV and indicate strong in-medium energy loss for charm quarks. The suppression is almost as large as that observed for charged (light-flavour) hadrons, with a possible
indication of $\raaD > \raah$, although not fully significant with the present level of experimental uncertainties.
The expected effect of nuclear shadowing is small (< 15\%) above $\pt$~= 6 GeV/c, indicating that the
large measured suppression cannot be explained by initial-state effects. Some of the pQCD models based
on various implementations of parton energy loss succeed reasonably well at describing simultaneously
the suppression of light flavour and charm hadrons. 
Single electrons and muons show similar suppression at and above $\pt >$ 5 GeV/$c$, where beauty decays are dominant according to NLO pQCD calculations. 
The precision of the D meson measurements will be improved in the future and the direct measurement of the $\raa$ of electrons from beauty decays should be possible using the large sample of Pb--Pb collisions recorded in 2011. 
In addition, the upcoming p--Pb run in 2012 will provide insight on cold nuclear matter effects in the low-momentum region.

\section*{Acknowledgments}
I would like to thank the organisers for inviting me and the stimulating atmosphere at this workshop. Bormio 2012 was a great experience for me.

The European Research Council has provided financial support under the European Community's Seventh Framework Programme (FP7/2007-2013) / ERC grant agreement no 210223.
This work was also supported by a Vidi grant from the Netherlands Organisation for Scientific Research (project number 680-47-232) and a Projectruimte grant from the Dutch Foundation for Fundamental Research (project number 10PR2884).

%


\begin{thebibliography}{99}

\bibitem{intro} Quark-Gluon Plasma 4, Eds. R.C. Hwa and X.-N. Wang, World Scientific Publishing, 2010.
\bibitem{deadcone1} Y.L. Dokshitzer, D.E. Kharzeev, Phys. Lett. {\bf B519}, 199 (2001).
\bibitem{deadcone2} M. Djordjevic, M. Gyulassy, S. Wicks, Phys. Rev. Lett. {\bf 94}, 112301 (2005).
\bibitem{theo:wicks2007} S. Wicks, W. Horowitz, M. Djordjevic, M. Gyulassy, Nucl. Phys. A784, 426 (2007).

\bibitem{alice:PPR2} P.~Cortese {\it et al.} [ALICE Collaboration], J. Phys. G: Nucl. Part. Phys. {\bf 32}, 1295 (2006).
\bibitem{alice:detpaper} K.~Aamodt {\it et al.} [ALICE Collaboration], J. Instrum. {\bf 3}, S08002 (2008).

\bibitem{alice:first1} J. Schukraft, arXiv:1112.0550.  
\bibitem{alice:first2} J.W. Harris (ALICE Collab.), Published by AIP in the Conf. Proc. Ser. (2011), arXiv:1111.4651.

\bibitem{alice:itsalign} K. Aamodt {\it et al.} [ALICE Collaboration], J. Instrum. {\bf 5}, P03003 (2010).
\bibitem{firstApaper} K.~Aamodt {\it et al.}, Phys. Rev. Lett. {\bf 105}, 252301 (2010). 
 
\bibitem{7TeVD} B.~Abelev {\it et al.} (ALICE Collaboration), JHEP {\bf 01}, 128 (2012).
\bibitem{RaaD} B.~Abelev {\it et al.} (ALICE Collaboration), submitted to JHEP {\bf } (arXiv:1203.2160), March 2012.
\bibitem{mu7TeV} B.~Abelev {\it et al.} (ALICE Collaboration), Phys. Lett. {\bf B708}, 265 (2012).

\bibitem{theo:fonll} M.~Cacciari {\it et al.}, JHEP {\bf 0103}, 006 (2001).
\bibitem{theo:vfns} B.A.~Kniehl, G.~Kramer, I.~Schienbein and H.~Spiesberger, arXiv:1202.0439.
B.A.~Kniehl {\it et al.}, Phys. Rev. Lett. {\bf 96}, 012001 (2006).

\bibitem{mnr} M. Mangano, P. Nason, G. Ridolfi, Nucl. Phys. B {\bf 373}, 295 (1992).

\bibitem{pionRaa} H. Appleshauser, J. Phys. G: Nucl. Part. Phys. {\bf 38}, 124014 (2011).
\bibitem{eps09} K.J. Eskola, H. Paukkunen and C.A. Salgado, JHEP {\bf 0904}, 065 (2009).

\bibitem{relB} D. Thomas, Hard Probes 2010. A. Mischke, J. Phys.: Conf. Ser. {\bf 316}, 012009 (2011). A. Mischke, Phys. Lett. {\bf B671}, 361 (2009).
 
\end{thebibliography}
\end{document}